# USID and Pycroscopy - Open frameworks for storing and analyzing spectroscopic and imaging data

Suhas Somnath[1], Chris R. Smith[2], Nouamane Laanait[3], Rama K. Vasudevan[2], Anton Ievlev[2], Alex Belianinov[2], Andrew R. Lupini[2], Mallikarjun Shankar[1], Sergei V. Kalinin[2], Stephen Jesse[2]

1. Advanced Data and Workflows Group, National Center for Computational Sciences
2. Center for Nanophase Materials Sciences
3. Computational Chemical and Materials Sciences, Computational Science and Engineering Division

Oak Ridge National Laboratory, Oak Ridge, TN 37831, USA

**Abstract**

Materials science is undergoing profound changes due to advances in characterization instrumentation that have resulted in an explosion of data in terms of volume, velocity, variety and complexity. Harnessing these data for scientific research requires an evolution of the associated computing and data infrastructure, bridging scientific instrumentation with super- and cloud-computing. Here, we describe Universal Spectroscopy and Imaging Data (USID), a data model capable of representing data from most common instruments, modalities, dimensionalities, and sizes. We pair this schema with the hierarchical data file format (HDF5) to maximize compatibility, exchangeability, traceability, and reproducibility. We discuss a family of community-driven, open-source, and free python software packages for storing, processing and visualizing data. The first is pyUSID which provides the tools to read and write USID HDF5 files in addition to a scalable framework for parallelizing data analysis. The second is Pycroscopy, which provides algorithms for scientific analysis of nanoscale imaging and spectroscopy modalities and is built on top of pyUSID and USID. The instrument-agnostic nature of USID facilitates the development of analysis code independent of instrumentation and task in Pycroscopy which in turn can bring scientific communities together and break down barriers in the age of open-science. The interested reader is encouraged to be a part of this ongoing community-driven effort to collectively accelerate materials research and discovery through the realms of big data.

# 1. Introduction

The rapidly decreasing cost of both computational power and storage in the past three decades, combined with advances in sensors and processing, has led to dramatic increases in the quality, quantity and complexity of data captured across a plethora of fields, ranging from weather satellites, personal healthcare devices, and social media to scientific imaging.

Here we focus on scanning transmission electron microscopy (STEM), which has emerged as one of the most powerful tools for material structure and functionality characterization at the atomic scale. The development of aberration correction[1-4], followed by its integration into commercial microscopes, has enabled the spread of routine atomic resolution imaging. Besides increasing the number of facilities producing such results, this development has increased the density and type of data available, for example, allowing spectroscopy from single atoms,[5] and extended the dimensionality of the data, such as from imaging to 3D focal series[6]. Other imaging modes such as 4D STEM, promise further increases in both spatial resolution and data dimensionality. [7-9] The data can be much more precise[10-12] and quantitative [13,14] allowing, subtle details, such as changes in shape, to be directly related to various physical properties [13,15-17]. Similarly, recent technical advances in monochromation[18-21] are facilitating investigations of physical phenomena associated with millivolt-level quasiparticles and excitations. Other novel spectroscopy techniques allow probing of magnetism and orbital physics in solids[18-21] at or near atomic resolution.[22-28] Finally, real-time feedback and control of the electron beam is being used for direct atom-by-atom fabrication[29], but produces a complex data-stream that might not have a traditional image shape[30].

The explosion in data volume, variety, and velocity resulting from the remarkable progress in instrumentation necessitate the development of equally advanced software, tools, and infrastructure to fully harness the potential in the data thereby seamlessly bridging data generation tools with modern high-performance computing (HPC) and cloud computing infrastructures[31-33].

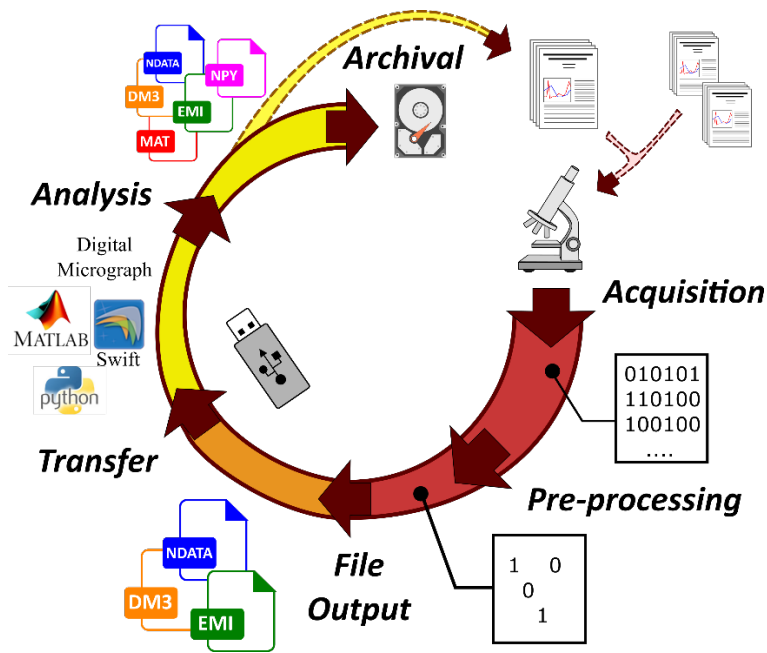

**Figure 1**. Current paradigm of research and lifecycle of datasets. Published literature informs the next experiment. The controller attached to the commercial instrument compress the information-rich data stream and writes the data into proprietary formatted files. A collection of proprietary and open-source software are used to analyze the data on personal computers. The raw data and results are all stored in conventional storage drives and the drive itself is physically archived once the researcher leave the research group.

From the data perspective, the factors that impede the progress of scientific research become evident when inspecting every stage of the lifecycle of datasets as shown in Figure 1. Currently, most STEM data is acquired from commercial instruments which tend to down-sample or discard the vast majority of the signal via averaging or decimation. As a consequence, crucial aspects of the sample properties are lost in the simplified signal. The down-sampled data are usually written into proprietary file formats, which impede and sometimes even preclude access to data and metadata, complicate long-term archival, obstruct sharing, and fracture the scientific communities along file formats.[34-36] The data are then transferred onto a personal computer using portable storage drives where the data is processed and analyzed using various combinations of proprietary and open-source software. The differences in structuring of data and metadata within the different proprietary files result in separate versions of similar codes for each modality or instrument vendor thereby impeding the sharing and reuse of code and data. The explosion in data volumes, due to continued development of instruments, is already rendering such modes of data transfer, conventional analysis software, and the use of personal computers[37-40] for computation infeasible. The results from the analyses are either written back to the original file in the proprietary format or into new files of other formats, thereby further expanding the number of file formats, data representation schemas, and software packages required to manage the data. Data are typically organized haphazardly and are not discoverable or searchable by metadata. Though findings from the data are reported in open literature, typically, neither the data nor the software are supplied with the journal article thereby thwarting open research and data mining. Finally, years of collected raw and processed data are forgotten as they remain on the personal computer which itself is physically archived when the researcher leaves the group (or upgrades to a newer computer).

All of these sub-optimal practices not only impede conventional modes of scientific discovery but also significantly thwart data-driven discovery, which requires large volumes of *well-curated* and standardized data in addition to robust and versatile software. Clearly, the technology, practices, and policies that underpin each stage of the data lifecycle need to be overhauled in order to support the needs of scientific discovery. In this article, we demonstrate a data model that can represent imaging and spectroscopic data of most common sizes, dimensionalities, scientific modalities, and instrument of origin. By combining such a general data schema with the most popular open file format for scientific data, we greatly simplify data access and pave the way for standardization of data analysis codes and scalable computing. We describe our user-friendly and open-source software packages that simplify access to data, while also taking advantage of cutting-edge algorithms and tools. Finally, we discuss how the practices and infrastructure underpinning other phases in the lifecycle of a data could be modified to facilitate data-driven scientific discovery in experimental and observational science facilities.

## 2. Data formats

Many commercially available scientific instruments generate data in proprietary file formats, with each manufacturer encoding the data and metadata in a different manner. Given the frequent need for custom analysis routines, access to data and metadata outside the manufacturer's software environment is a must. However, proprietary file formats make it either challenging or impossible to guess how the information in the file is structured, thereby significantly impeding data analysis.[34,41] Furthermore, comparison and correlation of data from multiple instruments and modalities becomes nearly impossible. Finally, the proprietary formats and software impede long-term archival and curation of data. Electron microscopes and similar other instruments are a large investment and tend to have a lifespan of several years or even decades. [42] Consequently, both the proprietary format and the software needed to access it need to survive the pace at which operating systems and computer architectures become obsolete. These problems are exacerbated for large laboratories and user facilities which need to manage software and data needs for dozens of instruments, while ensuring the efficiency of imaging workflows and allowing for data searchability and discoverability.

The solution for proprietary data formats can be broken into three components. The first is the data model, which provides an abstract representation of the data layout. The second is file format, which is the container in which the actual data is stored; such as text files (.txt), spreadsheets (.xlsx), image files (.png, .tiff, .jpeg), etc. The third is the combination of a schema and vocabulary that would represent the metadata. The first two components are loosely coupled since a given file format may not accommodate some data models from a fundamental or a user-friendliness perspective. For example, one could store image data in a spreadsheet, but it would not be convenient. While the first two components are technical challenges, the third is a sociological challenge since a solution necessitates quorum from the entire microscopy community.[43] We structure metadata in a hierarchical manner with broad top-level topics such as instrument, sample, and user. Each parameter in the metadata can be stored as key-value pairs such as "Instrument-Acquisition-Data_Sampling_Rate" = "100000 [Hz]". The units for physical values may need to be captured in a manner that is both human- and machine-readable.

Our goal was to develop a general and versatile model for representing data, regardless of instrument of origin, modality, and dimensionality. A generalized representation enables the development of instrument- and modality- agnostic data processing and visualization algorithms. For example, a singular value decomposition[44] or a clustering[45] algorithm should work regardless whether it is STEM, atomic force microscopy, or mass spectrometry data. This generality imposes some restrictions in the way data is laid out. Besides conventional data, we wanted to also accommodate edge cases where the data do not have an N-dimensional form. Some examples include compressed sensing[46], sparse sampling[47], or incomplete measurements where observations are not available for all combinations of the position parameters. Similarly, in certain cases, observations cannot be recorded for all combinations of spectroscopic parameters such as temperature, bias, etc. While the data could certainly be padded (with zeros or other default values) to create an N-dimensional form in such cases, doing so may change the meaning of the data and unnecessarily balloon the size of the dataset.

We evaluated existing scientific data exchange models within and beyond the STEM community such as NeXus[48], PaNData[49], HMSA[36], SPMML[35], MatML[50], Data Exchange[51], Open Microscopy Environment's (OME) version of the Tagged Image File Format (TIFF) – OME-TIFF[52], Electron Microscopy Data (EMD)[53], multidimensional eXtensible archive (MXA)[52,54], and the model that used by DREAM.3D[55], etc. Most of these were rigidly designed around specific kinds of instruments, modalities, coordinate systems (Cartesian only with no option for polar coordinates etc.), or scientific communities such as the climate research, X-ray microscopy, etc. Some models are tied to plain-text-based file formats that are limiting, especially for large multidimensional datasets[35,50]. While most of the models tied to binary file formats can certainly represent the vast majority of data, none besides HMSA could represent large datasets without an N-dimensional form. Furthermore, most models assume that experimental parameters vary linearly for all dimensions, thereby precluding modalities where a parameter, such as bias, was varied as a sine or a bipolar triangle.[56] The most general model as of this writing was the HMSA. However, its weakness was that a potential loss or separation of the metadata file from the data file would render the dataset useless.[54] To accommodate all, including edge cases, we developed a data model called Universal Spectroscopy and Imaging Data (USID). Details available at - https://pycroscopy.github.io/USID/about.html

## 2.1 Universal Spectroscopy and Imaging Data (USID)

The data in USID is stored in two main kinds of datasets: (i) *Main Datasets* that contain the actual data (measurement, analysis results, etc*.)* (ii) *Ancillary Datasets* that provide supplemental information regarding the *Main Dataset*.

### 2.1.1 Main Datasets

USID represents the measured data in a two-dimensional grid or a second order tensor called a *Main Dataset*. In these *Main Datasets*, the locations at which the measurements were collected are arranged along the first (vertical) axis and the observations (measurements) collected at each location are arranged along the second (horizontal) axis. All *Position* dimensions / parameters (e.g. – rows, columns, height, etc.) are collapsed to the vertical axis in the same way that all *Spectroscopic* dimensions (e.g. – voltage, temperature, pressure, etc.) are collapsed into the horizontal axis. The data is arranged in the same manner that reflects the sequence in which the data was collected. The *Main Dataset* is always associated with two mandatory attributes such as the physical *quantity* that was recorded in each observation along with the physical *units* of the quantity. Besides accommodating the edge cases, *Main Dataset* are already in the format (*instance* x *features*) accepted by machine learning and other processing algorithms.

**2.1.2 Ancillary Datasets**

*Main Datasets* are always accompanied by four *Ancillary Datasets* which are the *Position Indices*, *Position Values*, *Spectroscopic Indices*, and *Spectroscopic Values*. The *Position Values* and *Spectroscopic Values* datasets are analogous to map legends. In other words, these two datasets provide the combination of the values for each dimension / independent parameter that correspond to a particular data point in the main dataset. The *Position Indices* and *Spectroscopic Indices* datasets are counters for each position and spectroscopic dimension / parameter. While the user may not directly derive much value from the *Indices* datasets, they are critical for slicing subsections of, reducing, and transforming the two dimensional *Main Dataset* into its original N-dimensional form, if available.

Like the *Main Datasets*, the *Ancillary Datasets* are always two-dimensional matrices regardless of the number of position or spectroscopic dimensions. Let us consider a *Main Dataset* collected over $N$ positions, each containing $S$ spectral values with $U$ position dimensions and $V$ spectroscopic dimensions. The size of the 2-dimensional matrix that makes up this *Main Dataset* would be ($N$ x $S$). Then, the *Position Indices* and *Position Values* datasets would both be of the same size – ($N$ x $U$). Similarly, the *Spectroscopic Values* and *Spectroscopic Indices* dataset would both have a size of ($V$ x $S$). The columns in the *Position Indices* and *Values* datasets and the rows in the *Spectroscopic Indices* and *Values* datasets would be arranged in ascending order of the rate of change. In other words, the first column in the *Position* datasets would be the fastest changing position dimension, while the last column would be the slowest changing position dimension. Therefore, each dimension would get its own row in the *Spectroscopic* datasets or a column in the *Position* ancillary datasets. Much like the *Main Datasets*, the *Ancillary Datasets* are also always associated with mandatory attributes for the names of all position or spectroscopic dimensions and the physical units of each dimension. Optional attributes would be used to indicate the names of incomplete dimensions (e.g. - incomplete measurement, sparse sampling) and dependent dimensions.

### 2.1.3 Examples

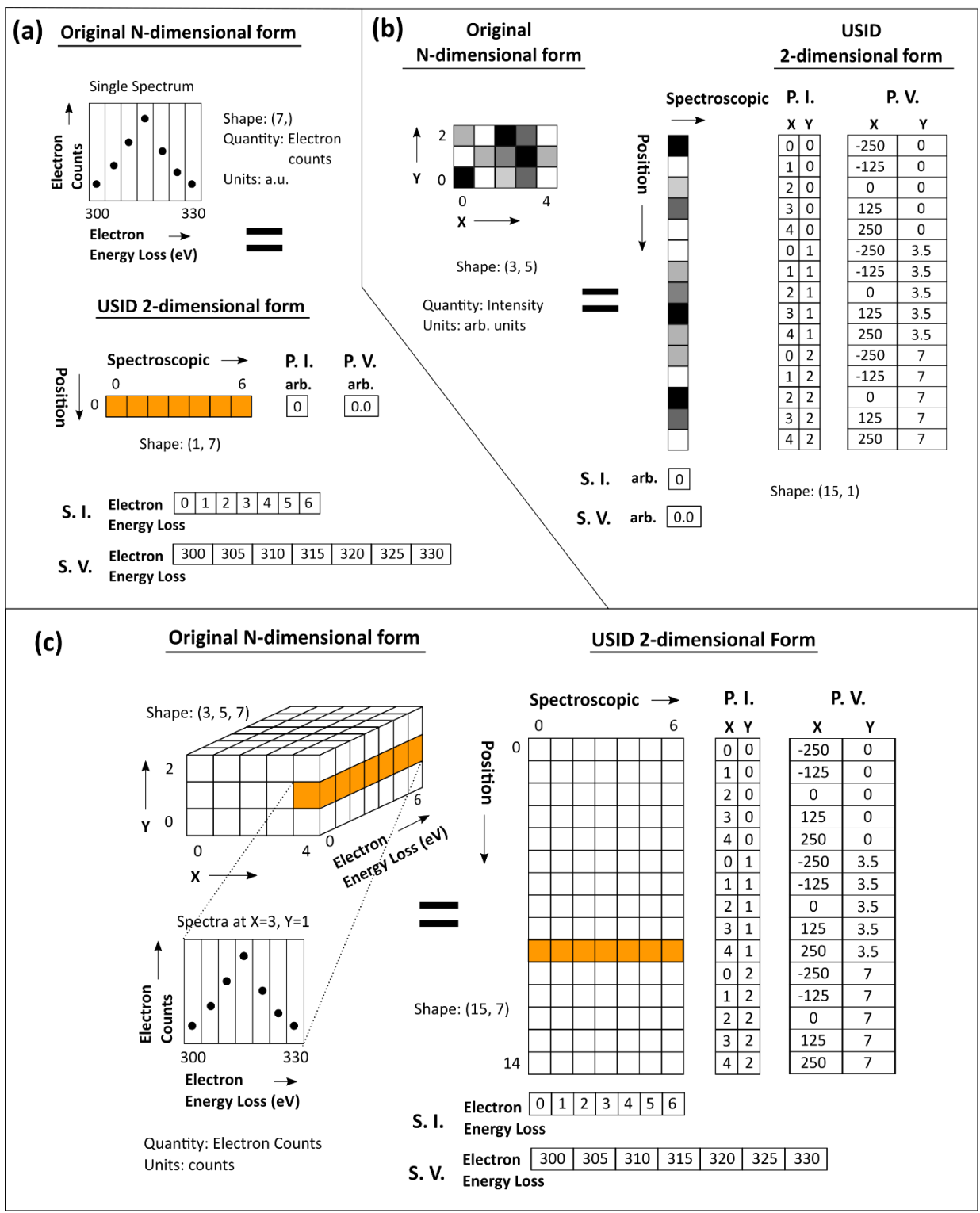

**Figure 2**. Schematics illustrating how data would be represented via the Universal Spectroscopy and Imaging Data (USID) model. The USID model stores observational data in a two-dimensional matrix called the 'Main' dataset wherein locations at which the measurements were collected are arranged along the first axis while the data collected at each location are laid out along the second axis. Each 'Main' dataset is accompanied by four ancillary datasets. The 'Position Values' (P.V.) and 'Spectroscopic Values' (S.V.) datasets explain the unique combination of parameters that result in each data point in the 'Main' dataset. The 'Position Indices' (P. I.) and 'Spectroscopic Indices' (S. I.) datasets serve as counters for each independent position or spectroscopic dimension and contain vital information regarding the original dimensionality of the 'Main' dataset. The USID representations are provided for (a) 1D spectrum, (b) 2D gray-scale image, and (c) 2D map of spectra.

Fig. 2 illustrates the USID representation for three simple cases.

### 2.1.3.1 Spectrum

The simplest example is a single spectrum, such as an electron energy loss spectroscopy (EELS) spectrum. Here, seven data points have been acquired as a function of seven values of a single parameter, *Electron Energy Loss* as shown in Fig. 2(a). Since it is a single spectrum, it has an arbitrarily defined *Position* dimension, *X* varied over a single value. The data was acquired as a function of a single *Spectroscopic* dimension - *Electron Energy Loss* and the data points are arranged column-wise in a single row. The USID size of such a *Main Dataset* is (*1* x *7*), where *1* signifies the single location of acquisition.

Consequently, the *Position Indices* and *Position Values* datasets both have a single column to represent the arbitrary position dimension and a single row to represent the single position of acquisition. Both *Position* datasets have a size of (*1* x *1*). The independent, *Spectroscopic* parameter - *Electron Energy Loss* was varied over seven values from *500* to *900 eV*, the *Spectroscopic Indices* and *Spectroscopic Values* datasets both have a single row for the *Electron Energy Loss* dimension and five columns for each of the seven values over which the parameter was varied. Therefore, the ancillary *Spectroscopic* datasets have a size of (*1* x *7*). The *Spectroscopic Values* dataset simply list the values over which the *Electron Energy Loss* was varied in the seven columns while the entries in the *Spectroscopic Indices* dataset ranges from *0* to *6* as a simple counter of the seven steps in the *Electron Energy Loss* dimension.

### 2.1.3.2 Gray-scale Image

In the case of gray-scale images such as annular dark or bright field (ADF or ABF) imaging, the quantity *Intensity*, is acquired as a function of two *Position* parameters, *X* and *Y* of size *5* and *3* respectively as shown in Fig. 2(b). In USID, the *Position* dimensions need to be collapsed along the first axis. Thus, the columns in the image would have to be arranged end-to-end along the first (vertical) axis, assuming that the data was collected column-by-column and then row-by-row. Given that only a single observation, or data point was collected at each location in the image, this dataset has a single arbitrary *Spectroscopic* dimension that could be named *arb*, which varied over a single value. Thus, the (*3* x *5*) shaped gray-scale image is represented as a (*15* x *1*) *Main Dataset* in USID.

This *Main Dataset* is accompanied by *Position Indices* and *Position Values* datasets of size - (*15* x *2*). The data for *X* was collected from *-250* to *+250 nm* for each *Y* from *0* to *7 μm* and *X* was varied faster than *Y*. Therefore, the values *[-250, -125, 0, 125, 250]* are repeated thrice along the first column of the *Position Values* dataset for dimension *X*. In the same way, the value *0.0* is repeated five times followed by *3.5* repeated five times followed by *7.0* repeated five times along the second column of the *Position Values* dataset. Since the *Position Indices* simply contain counters for each dimension, the first column contains [0, 1, 2, 3, 4] repeated thrice for dimension *X*. In the same way, the second column simply repeats *0*, *1*, and *2* five times each for dimension *Y* as shown in the figure. Notice, that the general nature of USID allows the dimensions to have different scales (*X* in nanometers and *Y* in microns). If one were interested in the data at the seventh position, the *Position Values* dataset would show that the data was acquired for the combination of *X* of *-125 nm* and *Y* of *3.5 μm*. The *Spectroscopic Indices* and *Values* datasets are sized (*1* x *1*) to accommodate the arbitrary spectroscopic dimension similar to the *Position* datasets for the spectrum example. Image distortions, drift correction, and transforms can be handled naturally in USID since the *Main Dataset* remains the same and changes are only applied to the *Position Values* dataset.

### 2.1.3.3 Spectral Maps

If a spectrum of length *7* was acquired as a function of a parameter, such as *Electron Energy Loss*, at each location on a two-dimensional grid of positions, it would produce a three-dimensional dataset as shown in fig. 2(c). Much like the grayscale image, this dataset has two *Position* dimensions - *X* and *Y* of sizes *5* and *3* respectively, but this dataset also has a single *Spectroscopic* dimension - *Electron Energy Loss* like the single spectrum example. Similar to the grayscale image, the *X* and *Y* dimensions are flattened along the first axis of the USID *Main Dataset* while the spectrum at each location is laid out along the second axis. Thus, the original (*3* x *5* x *7*) 3D dataset would be is flattened into a (*15* x *7*) USID *Main Dataset*. Since this dataset has the same *Position* dimensionality as the gray-scale image example, the *Position Indices* and *Position Values* datasets for the two examples are the same. Since this dataset has the same *Spectroscopic* dimension as the single spectrum, the *Spectroscopic Indices* and *Spectroscopic Values* datasets for the two examples are the same.

### 2.1.3.4 Other

Additional examples, including a few hands-on examples of real data in USID, such as time-series or 'movies', are available at: https://pycroscopy.github.io/USID/auto_examples/index.html. Similar to measured data, derived datasets such as results from analyses or processing algorithms can also be represented in USID. *Main Dataset* or *Ancillary Values Datasets* can accommodate real-value, complex-value, and compound (e.g. - red, green, blue components of color image) data without modifications. The USID model also accommodates data acquired *simultaneously* from separate information channels such as the annular bright-field and dark-field sensors in STEM. USID has successfully represented data from dozens of instruments, experiment modalities, and dimensions ranging from 1 to 8. If completely different measurements are recorded at each location, other data models or even data storage paradigms (such as databases) may be better suited compared to USID.

### 2.2 File Formats

By its definition, USID is an abstract data model and not accessible or tangible without its implementation into a file format. Therefore, we searched for file formats that would meet the following requirements: (a) established standard in scientific research; (b) accommodates multiple datasets of different shapes, dimensionalities, precision and sizes; (c) accommodates metadata; (d) scalable from a few kilobytes to several terabytes; (e) intuitive in structure, organization, installation (if required) and usage; (f) capable of storing relationships between datasets; (g) allows compression and decompression of data; (h) accessible without[57] and with any programming language including Matlab[58], R, python[59], C++[60,61], C#[62], Java[63] and perl[64], Igor Pro, etc.; and finally (i) compatible with cloud and high-performance computing by supporting parallel read and write[65]

Among the many file formats we tested including MathWorks' .MAT files[58], XML[66] OME-TIFF files[52], home-made binary files, common data format (CDF)[67], net common data format (NetCDF)[68], hierarchical data format (HDF5)[69], Zarr[70], etc., HDF5 was the only file format that satisfied all our requirements. In fact, HDF5 supports most popular data models such as NetCDF, NeXus[48], Coherent X-Ray Imaging (CXI) data[71], DREAM.3D model[55], and even MathWorks's .MAT. HDF5 can be thought of as a 'file-system within a file.' HDF5 *Datasets* are analogous to files and can contain multidimensional data arrays while *Groups* are analogous to folders and can contain a collection of groups and/or datasets.

The HDF5 implementation of USID, referred to as h5USID, implements USID *Main* and *Ancillary Datasets* as HDF5 *Datasets*. HDF5 *Groups* are used to organize data acquired from different detectors and results from data analyses. HDF5 *Attributes* are used to store metadata such as 'quantity', 'units', etc. in HDF5 *Groups* or *Datasets*. Different signals or physical quantities acquired simultaneously (e.g. – annular bright-field and dark-field detectors in STEM) can be accommodated under separate HDF5 *Groups* named as *Channels* in h5USID files. Complete specifications for h5USID are available at https://pycroscopy.github.io/USID/h5_usid.html.

Traceability and reproducibility of the experiments as well as the data processing routines in h5USID are captured through comprehensive and consistent metadata stored as attributes, in addition to links created between the data source and the analysis results within the HDF5 file. Keeping the metadata and data within a single file, h5USID mitigates the need for a traditional relational database for tracking data processing via frameworks such as Sumatra[72], ActivePapers[73], or Madagascar[74], though they can be added on if desired.[75] In a subsequent update to h5USID, we plan on exploring advanced use-cases where results of data processing are written into independent files, tracking provenance across multiple files, data processing on datasets present in multiple files, etc. h5USID already satisfies most of the requirements of FAIR principles[76] and we intend to satisfy all the recommendations put forward. Data stored in h5USID files is curation-ready and therefore meet the guidelines for data sharing issued to federally funded agencies as outlined by the United States Department of Energy.[77]

## 3. Software

The software provided with (commercial) instruments is typically proprietary or black-box in nature, which usually either precludes or significantly impedes the automation of and improvements to experiments and data acquisition.[78] Furthermore, such software packages are often inadequately equipped for advanced data analysis routines, typically do not allow users to add or edit functionality, are expensive to license and in certain cases even require hardware encryption keys.[79] In an age where experiments are becoming so complex that mere textual representation in a paper is woefully inadequate, these proprietary software exacerbate the problem, since they were not designed with traceability and reproducibility of experiments or data sharing (as now progressively being required by journals and funding organizations) in mind.[75] As a consequence, many researchers typically resort to third-party software for analyzing their data.

Though third-party data analysis software offer many important features, a few key problems remain unaddressed. First, most third-party software only read data in proprietary file formats and then perform data processing and visualization. None write results into standardized file formats, let alone develop a community standard for the data model. Second, most software are limited to analyses on simple one, two, and three-dimensional data of small sizes. Almost no software support operations on large (1 GB - 1 TB) and multidimensional datasets or allows seamless scaling of computations from personal computers to HPC/cloud resources. Third, most analysis routines provided by the software are often simple and straightforward and they do not offer advanced statistical analysis or machine learning capabilities. Fourth, while many open-source software allow anyone to modify the code to suit their purposes, the ability for the average user to do so is limited by the quality of the documentation provided by developers[41,78] and the difficulty of the programming language (for example, writing code in C / C++[60,61] or Pascal is noticeably harder than doing so in Python[59] or Matlab[58]). As a consequence, it may be challenging to adopt such software, especially if the developers(s) have stopped all development or provided insufficient documentation. This is especially true of scientific software which are largely written by graduate students and postdoctoral researchers who rarely continue software development beyond their projects in academia. Fifth, many software are not laid out in an intuitive and modular enough manner to allow users to extend the software to their liking or scale the software to distribute computing over a cluster of computers. Finally, some of the available software are closed-source and therefore the algorithms used to process the data are opaque to the researcher.

We considered several programming languages such as Matlab[58], R, python[59], C / C++[60,61], C#[62], Java[63] and Perl[64] when choosing the language for our software packages but we finally chose python for a number of reasons.[80] First and perhaps most importantly, the user-friendliness and high-level nature of python significantly lowers the barrier for researchers to learn programming. Users can find answers to popular and even niche questions on forums.[81] Second, the Python ecosystem is rich in a variety of add-on packages for efficient numerical computing[82-84], image processing[85,86], scientific analysis[87], machine-learning[88], deep-learning[89-91], file-handling[92], graphical user interfaces (GUI)[93], graphical processing unit (GPU) computing[94], parallel computing[95-98], and data visualization [99,100] that dramatically simplify and shrink the time necessary to develop code.

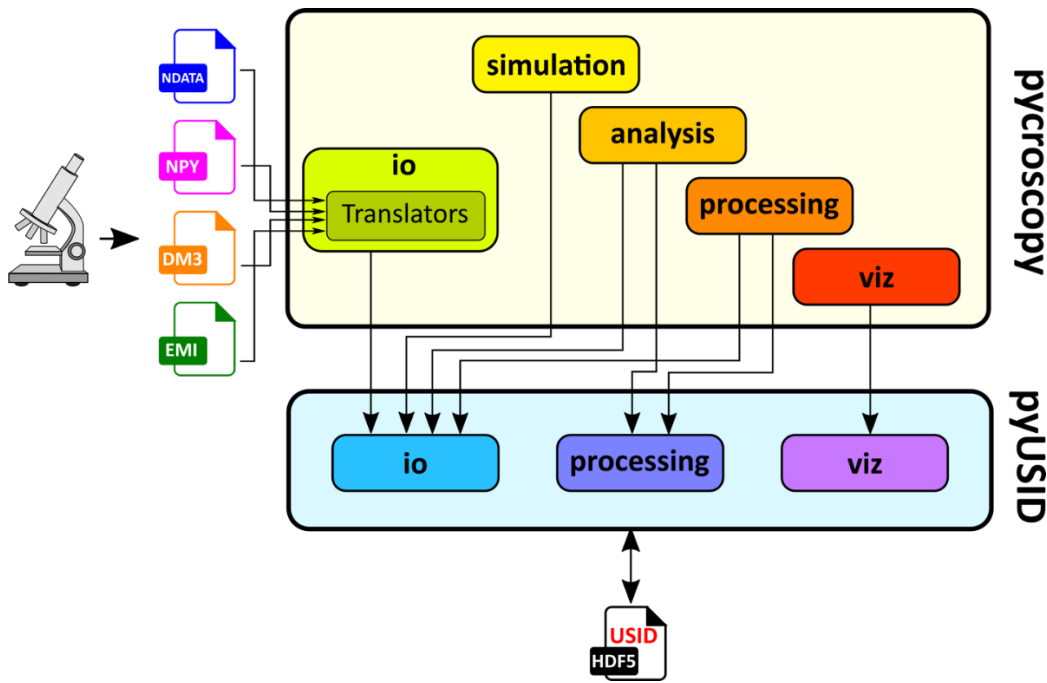

**Figure 3**. Interconnectivity in the USID ecosystem. Proprietary files generated by instruments are transformed to USID HDF5 file via Translator classes in the io subpackage of Pycroscopy. Once the USID HDF5 file is generated, the user has access to numerous analysis, visualization, and processing routines in Pycroscopy. All the file operations, data processing, and visualization operations in Pycroscopy use the *io*, *processing*, and *viz* sub-packages within pyUSID.

Third, the majority of the existing packages for material science and microscopy, especially for reading proprietary file formats are sometimes only available in python. Finally, unlike software like Matlab or IDL[101], which are prohibitively expensive for use outside academia, python and nearly all relevant packages are freely available.

We have developed a science-agnostic / engineering-fundamentals python package called pyUSID that supports an imaging and spectroscopy focused, pure-science package called Pycroscopy. The interactions between these packages is shown in Fig. 3. Both software were developed by and for microscopists, material scientists, physicists, etc., who need and use the software instead of employing dedicated software programmers.

**3.1 PyUSID**

pyUSID is the python interface for USID files, specifically h5USID files. Its structure has been divided into three main sub-packages - *io*, *processing*, and *viz*, along the main types of functionalities it provides. The *io* sub-package is the primary component of pyUSID and provides several utilities and python classes that simplify reading and writing h5USID files. The *ImageTranslator* and *NumpyTranslator* classes convert conventional image files or any data in memory into USID datasets in h5USID files. Once the data is translated to h5USID files, the functionalities in the rest of pyUSID and other USID-based scientific packages such as Pycroscopy become readily accessible. The *USIDataset* class gathers information from the ancillary datasets to simplify slicing, visualizing, reducing, and reshaping the flattened USID dataset to its original N-dimensional form. pyUSID also provides tools to handle transformation of complex and compound-valued datasets to real values and vice versa for compatibility with statistical analysis algorithms, which only work on real-valued data.

The *Process* class, provides an application-agnostic framework for formulating scientific and analytics problems into computational problems and enables piecewise and parallel processing of arbitrarily large datasets. Only the instructions for reading, writing the data in the h5USID file and the computation that will be performed on data corresponding to a single location data in the case of inherently parallel problems such as functional fitting, signal filtering, etc. needs to be provided. The *Process* class handles routine operations including memory management and parallelization of the computation over all available cores in a single CPU or multiple CPUs in a cluster of computers in cloud and high-performance-computing environments to significantly reduce the data processing time. The same scientific application written for personal computers will work on clusters. Since the *Process* class stores all necessary metadata within the h5USID file, it is also capable of checking the h5USID file for existing results from a requested computation and resuming partially-completed computations on similar or different computer architectures.

Finally, the *viz* sub-package of pyUSID has several plotting functions that simplify the visualization of data and generation of figures for journal publications. In addition, *viz* also has interactive widgets that greatly simplify the exploration of multidimensional data. pyUSID uses a strong defensive style of programming with comprehensive validation of inputs to functions with descriptive error messages in an effort to avoid accidental damage to the data files.

Empirically, the appeal and reach of a software is limited by the quality of the documentation.[78,102] Towards this end, pyUSID provides examples for every function to guide users to not only understand and use pyUSID, but also start developing code for and using pyUSID without additional help. Additional information on pyUSID is available at https://pycroscopy.github.io/pyUSID/about.html

## 3.2 Pycroscopy

Pycroscopy is a python package built on pyUSID for the scientific analysis of nanoscale imaging and spectroscopy modalities such as multi-frequency scanning probe microscopy, scanning tunneling spectroscopy, x-ray diffraction microscopy, and transmission electron microscopy. Pycroscopy consists of five main sub-packages:

- *io*: *Translator* classes that translate data from various proprietary file formats to h5USID files.
- *processing*: physical-model-independent processing of data including statistical learning, image processing, and signal filtering.
- *analysis*: physical-model based (inverse) analysis of data
- *simulation*: (forward) modeling of physical phenomena
- *viz*: interactive widgets and plotting tools specific to classes in *processing*, *analysis*, and *simulation*.

The *processing* subpackage of Pycroscopy provides file reading and writing wrappers for popular machine learning functions in packages such as scikit-learn via the *Cluster*[45], Singular Value Decomposition[44] - *SVD*, and *Decomposition* classes. These classes automatically handle the conversion of compound or complex data to real data and vice versa since most statistical analysis algorithms either do not accept or provide non-physical results for complex- or compound-valued data.

At the moment, Pycroscopy has translators for files from:
- STEMs: Digital Micrograph, Nion Company, OneView cameras,
- AFMs: Asylum Research, Bruker, and Molecular Vista
- STMs: Omicron
- SPM - general: Nanonis controllers, Gwyddion[78]
- ORNL developed modalities: Band excitation[103,104] and General-mode[105-107] related modalities SPM modalities.

The rest of the utilities in Pycroscopy become accessible once the h5USID file is obtained after translation.

Scientific workflows in Pycroscopy are disseminated via Jupyter notebooks[108], which are interactive documents that contain documentation, equations, images, code snippets, plots and interactive widgets[109]. Such self-explanatory notebooks can serve as a powerful narrative tool to demonstrate how the results were derived from the raw measurement data. Jupyter notebooks using Pycroscopy are now regularly submitted as supplementary documents along with the conventional manuscript to academic journals to improve transparency in the push for open science. Beyond Jupyter notebooks, we enhance reproducibility in data analysis by packaging the aforementioned notebooks, necessary software packages, and the data into a portable Singularity[110] or Docker[111] container that can be run anywhere at any time.

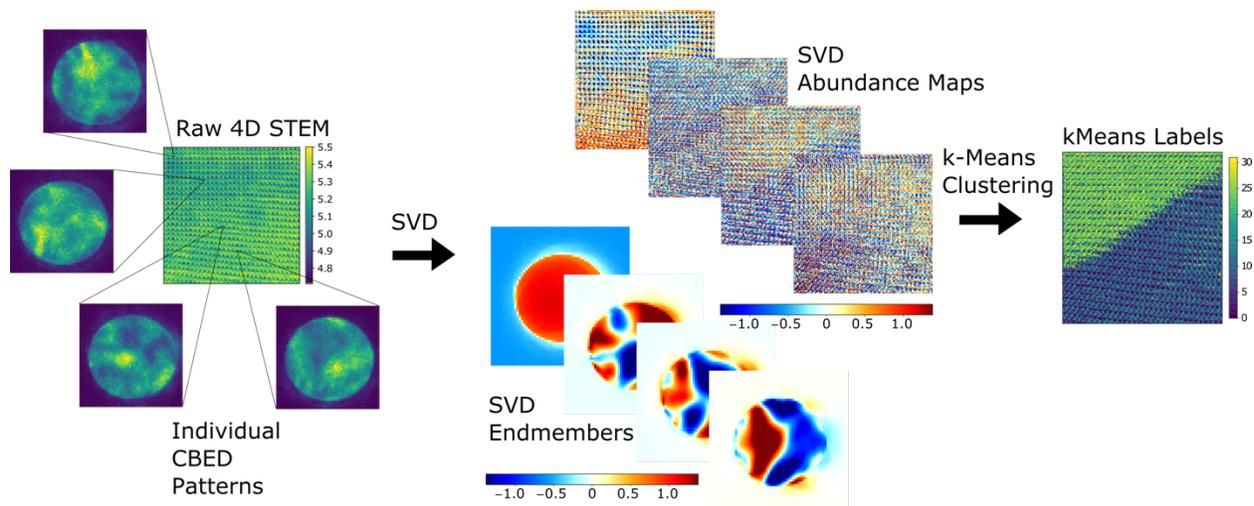

**Figure 4**. Using multivariate statistics to extract hidden trends in 4D STEM data.[112] The conventional HAADF image of the sample (left) does not reveal some of the most interesting physical features. However, applying a combination of singular value decomposition (SVD) and clustering to the complete 4D STEM dataset reveals a phase boundary in the data (right) that was almost invisible within the HAADF image.

Figures 4-5 shows two examples of recent scientific accomplishments enabled by Pycroscopy. Fig. 4 shows the use of the *SVD* and *Cluster* classes in Pycroscopy to extract trends in 4-D STEM datasets that are invisible to the human eye.[112] Typically, the raw Convergent Beam Electron Diffraction (CBED) patterns (2D images) recorded by the camera detector are averaged at each position of the electron beam to provide a pair of 'bright-field' and 'dark-field' 2D images. Performing SVD on the complete 4D dataset and retaining only the first few components captures the most statistically significant trends. The labels obtained by performing k-means clustering on the abundance maps produced by SVD clearly reveal a domain boundary that is invisible in the bright-field or dark-field images. This type of analysis highlights the data driven scientific paradigm, where the improvements in instrumentation and the detectors push the technique to extend beyond its classical capabilities. Traditionally STEM is not sensitive to ferroelectric domain boundaries, however the combination of novel hardware technology and a fresh perspective on analysis enhance the technique beyond its traditional characterization niche.

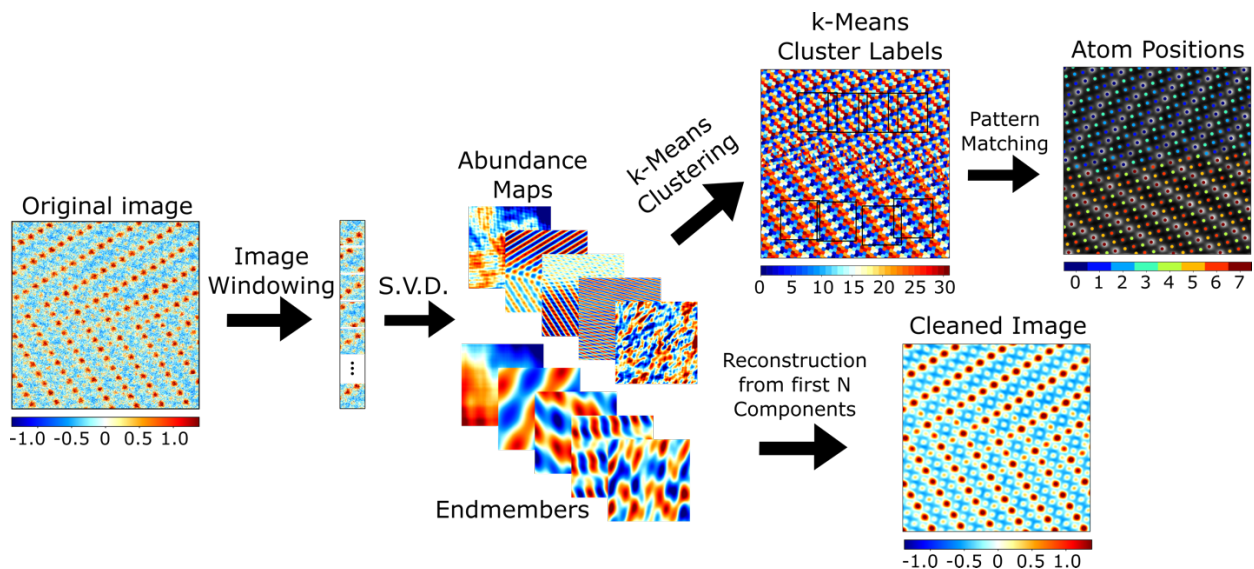

**Figure 5**. Statistical image cleaning and atom finding via Pycroscopy.[113] A stack of sub-sections of the original image was created by sliding a small window along the image. Singular value decomposition (SVD) was applied to the stack of image sub-sections to decompose the data into components with decreasing statistical significance. Reconstructing the data with only the first few (statistically most relevant) components resulted in an image with significantly reduced noise. The positions of different atomic columns in the image were semi-automatically extracted by applying clustering to the first few SVD components and using pattern-matching techniques.

Fig. 5 is an example where statistical methods significantly lowered the noise in atomically resolved images and found atomic positions.[113] The *ImageWindowing* class in Pycroscopy was used to build a stack of sub-images by sliding a window across the image.[114] The same generalized *SVD* class used in the previous example was used again to decompose the data into principal components arranged in descending order of statistical significance. Reconstructing the data only using the first few physically (and statistically) relevant components resulted in a similar image with substantially reduced noise. The same *Cluster* class from above was used for k-Means clustering on the abundance maps from SVD to reveal similarities between different sections in the image. Once the user identified repeating patterns, that correspond to different atomic columns, algorithms in Pycroscopy were not only able to pin-point the locations of atoms but also differentiate the different atoms and the same atoms belonging to different phases in the material. Additional information on Pycroscopy is available at https://pycroscopy.github.io/pycroscopy/about.html.

Both Pycroscopy and pyUSID can be used wherever python can be installed on a variety of operating systems such as Windows, MacOS, and Linux. As of this writing, pyUSID and Pycroscopy contain in excess of 41,700 (non-blank) lines of code and 3,500 (non-blank) lines of documentation. Pycroscopy has been downloaded in excess of 130,000 times[115] while pyUSID has been downloaded over 4,500 times[116]. The source code for both pyUSID and Pycroscopy has been developed in a transparent and traceable manner using the Git[117] version control system.[118] We have made every effort to ensure that our code is readable to the average material scientist by providing comments everywhere appropriate. In addition, both packages use multiple continuous integration (CI)[119] services, such as Travis CI[120], CircleCI[121], and AppVeyor[122], that automatically build the software and documentation in addition to catching and reporting errors in the code and documentation for multiple combinations of operating systems, python versions, and dependency package versions. Both the packages are available through the most popular distribution sources for python software such as the python package index (pypi) and Conda Forge. The packages are installed easily and in a consistent manner using a simple command as 'pip install pycroscopy'.

Neither pyUSID nor Pycroscopy are finished products; rather they are constantly evolving based on the requirements and contributions of the users from around the world. The interested reader is encouraged to be a part of this community-driven effort by engaging with the user and developer communities on our forums, submitting their own algorithms, data, extractors to read proprietary data files, pointing out or fixing bugs, requesting new features, or guiding the development of future features. Users are also encouraged to discuss topics on public forums for pyUSID (https://groups.google.com/forum/#!forum/pyusid) and pycroscopy (https://groups.google.com/forum/#!forum/pycroscopy). Besides offering USID, pyUSID, and Pycroscopy, we engage with user-community through online discussion forums. In addition, we lower the barrier to advanced data analytics by routinely organizing workshops[123-127] to educate the microscopy and the materials science communities on a broad range of topics ranging from basic python and best practices for scientific programming[128] to image and spectral analysis to machine learning and deep learning applied to material science.

## 4. Discussion

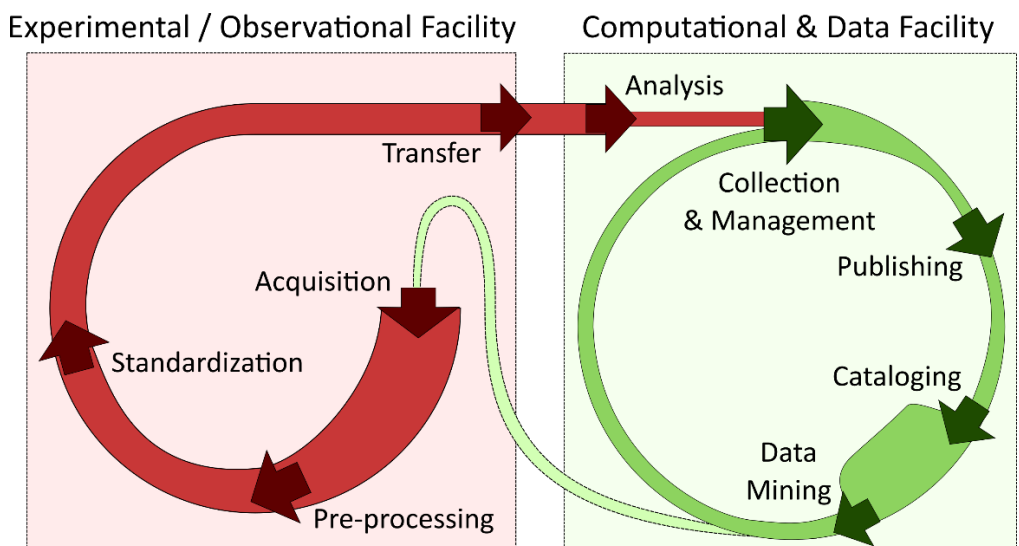

**Figure 6**. Lifecycle of datasets for the proposed future paradigm of research. The complete raw-data stream is acquired from the instruments and written to open and standardized data formats. The raw data is transferred to a partner computational facility and rapidly analyzed on a cluster of computers. Subsequently, the raw data and results are stored methodically. Important datasets are published and made available through data catalogs. Data catalogs facilitate data driven scientific discovery and inform the next experiment.

As mentioned earlier, careful analysis and retooling of the infrastructure, policies, and practices at each step in the lifecycle of datasets is necessary for facilitating data driven discovery in the era of big data. This article discussed our solutions for storing and analyzing scientific imaging and spectroscopy data. Here we briefly discuss suggestions for the rest of the components in the data lifecycle as shown in fig. 6.

High throughput experimentation can be enabled by automating portions of the experiments by facilitating streaming, real-time analysis, and feedback of measured data from instruments.[105,107] We realize that this is a challenge in STEM and other experimental sciences where commercial instruments and proprietary instrument-control software impede such automation. Nonetheless, many instruments do allow access to record and retain the complete signal before heterodyne-like detection systems down-sample the signal. Prior demonstrations of General-mode[105-107] have shown that the availability of the complete signal provides dramatic improvements to the measurement speed and detection limits in addition to being unbiased. Once the measured data and metadata are written into standardized files, the data should be transferred rapidly and reliably to a storage repository via the best possible data transfer standard.[129] Research data should be stored in robust, failure-safe file storage systems, either at a partner data / computational facility (if available) or on the cloud; with tiered remote access if desired. The large volumes of data generated by multiple instruments and used by multiple researchers collaboratively also necessitates a scientific data management software capable of searching, discovery, organization, dissemination, fine-grain access control and sharing, movement across data and computational facilities, etc.[130]

Once the data is well managed, the analysis should be driven by open-source, community-developed software that scale with the size of the data as well as computational resources. The barrier to the data and computational resources can be lowered through a collaborative and interactive analytics environment[131]. Relevant datasets could be peer-reviewed and awarded a digital object identifier (DOI) number and added to a catalog[132-134] which could be mined using machine learning and deep learning techniques to learn broad trends. Components such as the data storage, publishing, cataloging, collaborative analysis environment, and the distributed computing resource would fall under the purview of a partner computational facility. Other components such as the optimization and scaling of scientific analysis software, transfer of data from instruments to the repository and data streaming from instruments would require a collaborative effort between experimental and computational groups, while the remaining components are mainly the responsibility of the domain-science researchers.[135] Eventually, all stages subsequent to data acquisition may be handled in computational facilities instead of the experimental facilities similar to how the synchrotron facilities operate today. While this paper focused on the STEM community, other domains such as SPM, X-ray microscopy, etc. share similar concerns. Solutions discussed here can be applied to many other scientific domains within and beyond microscopy.

We feel that it is the responsibility of the community (researchers, industry, funding agencies, scientific journals, etc.) to collectively define data structuring and file format standards. After all, it is the absence of these community standards that brought about the proprietary format dissonance. Developing standards is not an impossible task; several scientific societies such as the climate research[68], geospatial research[136], astronomy[137], high energy physics[48], biological and biomedical imaging[52] communities have long established standards. The use of such community standards can be enforced by working with scientific journals to mandate that data supporting an article be submitted in the community standard. Similarly, federal funding agencies could insist that instruments purchased with federal funds export data to community standards and completely reveal the structure of proprietary formatted files.

In the modern STEM data landscape, we anticipate researchers choosing to work with a collection of analysis software. While it is possible to draw bridges between software efforts, this is typically a tedious process and can be especially challenging if the software are written in different languages. We believe that it is substantially easier for software to exchange information by being compatible with common data schemas such as USID and a file format like HDF5. The strong documentation on USID in combination with user-friendly functions in pyUSID can enable new as well as existing scientific python packages, such as HyperSpy[138] or LiberTEM[139], to exchange data with other packages such as Pycroscopy by adopting the USID model and h5USID more readily.

In certain cases, such as in user facilities in national laboratories, a set of computationally intensive analysis or processing procedure needs to be performed routinely. Such situations necessitate and can justify the development of custom software that dramatically shrinks the time to solution. A subset of the authors in this paper recently demonstrated a cross-facility software that connected microscopes to supercomputers at Oak Ridge National Laboratory to enable near real-time analysis of large STEM and SPM measurement data using a convenient push-button interface.[140] However, the sheer complexity of the undertaking meant that making changes to the software was beyond the reach of the average material scientist. Lessons and findings from this project motivated the authors towards a two-pronged solution for software wherein the aforementioned streamlined HPC software would be complemented by a closely-mirrored, community-driven software developed by microscopists that prioritized simplicity and rapid development over outright efficiency. The most commonly used and computationally expensive algorithms in the latter would be optimized and incorporated in the former effort.

Since its inception, several researchers have started using the USID and h5USID in other scientific domains such as nuclear science, mass spectrometry, and in synchrotron facilities due to the simple universality of this data model. Following how pyUSID has helped the microscopy community develop a community-driven software solution through Pycroscopy, pyUSID could either spark the development of new crowd-sourced software packages in other communities or serve as a backend in existing community-driven software packages.

The ability of pyUSID's Process class to scale computations to clusters will be critical especially for new modalities such as multidimensional and high-frame-rate 4-dimensional STEM experiments that can generate datasets upwards of 500 GB - 10 TB. Such scalable computing framework will be especially beneficial for universities and national laboratories that host user facilities where the goal is typically to minimize the time spent on processing data and maximize time acquiring high quality data from instruments. We plan on gradually replacing computationally intensive portions of conventional python code with Cython, Numba[141], C++, or graphics processing units (GPU)-accelerated code to maximize performance. We also plan on encouraging researchers to contribute to software packages by adding infrastructure that prompts users of algorithms to cite the papers where the algorithm was reported.[41]

## 5. Conclusions

In this paper, we discussed the many technical and societal challenges that impede observational and data-driven scientific discovery. Accelerating scientific discovery necessitates retooling each component along the lifecycle of a scientific dataset. We presented the Universal Spectroscopy and Imaging Data (USID) model capable of representing most commonly observed data regardless of its size, dimensionality or lack of N-dimensional form, modality, precision, or even instrument of origin. We pair USID with the hierarchical data format (HDF5) file format, which was designed specifically to accommodate very large and multidimensional scientific data, operation on high-performance or cloud-computing infrastructure, etc. We also presented two community-driven, free, open-source, and user-friendly python software packages. The first of these is pyUSID that provides tools to read and write USID data in HDF5 files in addition to utilities for visualizing and scalable processing of data. The second package is Pycroscopy, which serves as a community-hub for sharing and deploying advanced algorithms for analyzing imaging and spectroscopy data from various nanoscale modalities. We will discuss our solutions for the other phases in the lifecycle of scientific datasets in subsequent articles.

**Authors' contributions**

AB and SJ conceived the original abstract structure of USID, while S.S. and C.R.S. implemented USID in code and developed all subsequent versions of USID. S.S. and C.R.S. are the cofounders of pyUSID. S.S., C.R.S. and N.L. cofounded Pycroscopy. S.S. and C.R.S. have been the core developers and maintainers of Pycroscopy and pyUSID since their inception. S.S., A.B., S.J, A.R.L., R.K.V., M.A. and S.V.K. prepared the manuscript. R.K.V. and A.I. made important contributions to popular scientific workflows in Pycroscopy. All authors read and approved the final manuscript.

**Acknowledgements**

This research was sponsored by the Division of Materials Sciences and Engineering, Basic Energy Sciences, US Department of Energy (S.V.K., A.R.L). This research was conducted and partially supported (C.R.S., R.K.V., I.A., S.S.) at the Center for Nanophase Materials Sciences, which is an US DOE Office of Science User Facility. A portion of this research was supported by the Laboratory Directed Research and Development Program of Oak Ridge National Laboratory, managed by UT-Battelle, LLC for the US Department of Energy (S.J., S.S., A.B.). This research used resources of the Oak Ridge Leadership Computing Facility (M.S., S.S.) at the Oak Ridge National Laboratory, which is supported by the Office of Science of the U.S. Department of Energy under Contract No. DE AC05 00R22725. Research performed by N.L. was supported by the Eugene P. Wigner Fellowship program at ORNL.

The authors thank the many contributors of Pycroscopy and pyUSID for providing feedback, suggestions, finding and fixing errors in the code and documentation. The complete list of contributors is available on the Pycroscopy and pyUSID repository pages on GitHub. The authors also thank Gerd J. Duscher for valuable feedback on the manuscript.

Notice: This manuscript has been authored by UT-Battelle, LLC, under Contract No. DE-AC0500OR22725 with the US Department of Energy. The United States Government retains and the publisher, by accepting the article for publication, acknowledges that the United States Government retains a non-exclusive, paid-up, irrevocable, world-wide license to publish or reproduce the published form of this manuscript, or allow others to do so, for the United States

Government purposes. The Department of Energy will provide public access to these results of federally sponsored research in accordance with the DOE Public Access Plan (http://energy.gov/downloads/doe-public-access-plan).

**Competing interests**

The authors declare that they have no competing interests.

**Availability of data and materials**

All the data and algorithms mentioned in this paper are freely available on the public software repositories – USID at [https://github.com/pycroscopy/USID](https://github.com/pycroscopy/USID), Pycroscopy at https://github.com/pycroscopy/pycroscopy and pyUSID at https://github.com/pycroscopy/pyUSID.

# References

1	Lupini, A. R., Krivanek, O. L., Dellby, N., Nellist, P. D. & Pennycook, S. J. in *Electron Microscopy and Analysis 2001  Institute of Physics Conference Series*   31-34 (2001).
2	Nellist, P. D. *et al.* Direct sub-angstrom imaging of a crystal lattice. *Science* **305**, 1741-1741 (2004).
3	Pennycook, S. J. & Nellist, P. D.    (Springer, New York, 2011).
4	Haider, M. *et al.* Electron microscopy image enhanced. *Nature* **392**, 768, doi:10.1038/33823 (1998).
5	Varela, M. *et al.* Spectroscopic imaging of single atoms within a bulk solid. *Physical Review Letters* **92**, 095502, doi:10.1103/PhysRevLett.92.095502 (2004).
6	Borisevich, A. Y., Lupini, A. R. & Pennycook, S. J. Depth sectioning with the aberration-corrected scanning transmission electron microscope. *Proceedings of the National Academy of Sciences of the United States of America* **103**, 3044-3048, doi:10.1073/pnas.0507105103 (2006).
7	Rodenburg, J. M. & Bates, R. H. T. The theory of super-resolution electron microscopy via Wigner-distribution deconvolution. *Philos. Trans. R. Soc. A Math. Phys. Eng. Sci.* **339**, 521-553 (1992).
8	Rodenburg, J. M., McCallum, B. C. & Nellist, P. D. Experimental tests on double-resolution coherent imaging via STEM. *Ultramicroscopy* **48**, 304-314 (1993).
9	Nellist, P. D. & Rodenburg, J. M. Beyond the Conventional Information Limit - the Relevant Coherence Function. *Ultramicroscopy* **54**, 61-74 (1994).
10	LeBeau, J. M., Findlay, S. D., Allen, L. J. & Stemmer, S. Quantitative atomic resolution scanning transmission electron microscopy. *Phys. Rev. Lett.* **100**, 206101, doi:10.1103/PhysRevLett.100.206101 (2008).
11	LeBeau, J. M., Findlay, S. D., Allen, L. J. & Stemmer, S. Position averaged convergent beam electron diffraction: Theory and applications. *Ultramicroscopy* **110**, 118-125 (2010).
12	Yankovich, A. B. *et al.* Picometre-precision analysis of scanning transmission electron microscopy images of platinum nanocatalysts. *Nat. Commun.* **5**, doi:4155 10.1038/ncomms5155 (2014).
13	He, Q. *et al.* Towards 3D Mapping of BO6 Octahedron Rotations at Perovskite Heterointerfaces, Unit Cell by Unit Cell. *ACS Nano* **9**, 8412-8419, doi:10.1021/acsnano.5b03232 (2015).
14	Jia, C. L. *et al.* Oxygen octahedron reconstruction in the SrTiO(3)/LaAlO(3) heterointerfaces investigated using aberration-corrected ultrahigh-resolution transmission electron microscopy. *Phys. Rev. B.* **79**, doi:081405 10.1103/PhysRevB.79.081405 (2009).
15	Borisevich, A. *et al.* Mapping Octahedral Tilts and Polarization Across a Domain Wall in BiFeO(3) from Z-Contrast Scanning Transmission Electron Microscopy Image Atomic Column Shape Analysis. *ACS Nano* **4**, 6071-6079, doi:10.1021/nn1011539 (2010).
16	Borisevich, A. *et al.* Exploring mesoscopic physics of vacancy-ordered systems through atomic scale observations of topological defects. *Phys. Rev. Lett.* **109**, 065702 (2012).
17	Li, Q. *et al.* Quantification of flexoelectricity in PbTiO 3/SrTiO 3 superlattice polar vortices using machine learning and phase-field modeling. *Nat Commun* **8**, 1468 (2017).
18	Krivanek, O. L., Lovejoy, T. C., Dellby, N. & Carpenter, R. W. Monochromated STEM with a 30 meV-wide, atom-sized electron probe. *Microscopy* **62**, 3-21 (2013).
19	Krivanek, O. L. *et al.* High-energy-resolution monochromator for aberration-corrected scanning transmission electron microscopy/electron energy-loss spectroscopy. *Philosophical Transactions of the Royal Society A: Mathematical,                       Physical and Engineering Sciences* **367**, 3683-3697, doi:10.1098/rsta.2009.0087 (2009).
20	Tiemeijer, P. C., Lin, J. H. A. & de Jong, A. F. First results of a monochromized 200 kV TEM, Microscopy and Microanalysis. *Microscopy and Microanalysis* **7 (sup. 2)** ( 2001).

21	Tiemeijer, P. C., van Lin, J. H. A., Freitag, B. H. & de Jong, A. F. Monochromized 200 kV (S)TEM. *Microsc. Microanal.* **8**, 70-71 (2012).
22	Schattschneider, P. Detection of magnetic circular dichroism using a transmission electron microscope. *Nature* **441**, doi:10.1038/nature04778 (2006).
23	Schattschneider, P. Detection of magnetic circular dichroism on the two-nanometer scale. *Phys. Rev. B* **78**, doi:10.1103/PhysRevB.78.104413 (2008).
24	Verbeeck, J., Tian, H. & Schattschneider, P. Production and application of electron vortex beams. *Nature* **467**, doi:10.1038/nature09366 (2010).
25	Idrobo, J. C. *et al.* Detecting magnetic ordering with atomic size electron probes. *Advanced Structural and Chemical Imaging* **2**, 5, doi:10.1186/s40679-016-0019-9 (2016).
26	Rusz, J., Bhowmick, S., Eriksson, M. & Karlsson, N. Scattering of electron vortex beams on a magnetic crystal: towards atomic-resolution magnetic measurements. *Phys. Rev. B.* **89**, doi:10.1103/PhysRevB.89.134428 (2014).
27	Rusz, J. & Idrobo, J. C. Aberrated electron probes for magnetic spectroscopy with atomic resolution: theory and practical aspects. *Phys. Rev. B.* **93**, doi:10.1103/PhysRevB.93.104420 (2016).
28	Rusz, J., Idrobo, J. C. & Bhowmick, S. Achieving atomic resolution magnetic dichroism by controlling the phase symmetry of an electron probe. *Phys. Rev. Lett.* **113**, doi:10.1103/PhysRevLett.113.145501 (2014).
29	Dyck, O. *et al.* Building Structures Atom by Atom via Electron Beam Manipulation. *Small* **0**, 1801771, doi:doi:10.1002/smll.201801771.
30	Sang, X. *et al.* Precision controlled atomic resolution scanning transmission electron microscopy using spiral scan pathways. *Scientific reports* **7**, 43585 (2017).
31	Kalinin, S. V. *et al.* Big, Deep, and Smart Data in Scanning Probe Microscopy. *ACS Nano*, 9068-9086, doi:10.1021/acsnano.6b04212 (2016).
32	Kalinin, S. V., Sumpter, B. G. & Archibald, R. K. Big-deep-smart data in imaging for guiding materials design. *Nat Mater* **14**, 973 (2015).
33	Sumpter, B. G., Vasudevan, R. K., Potok, T. & Kalinin, S. V. A bridge for accelerating materials by design. *NPJ Computational Materials* **1**, 15008 (2015).
34	Barrett, S. in *Proceedings of the Royal Microscopical Society.* 167-174 (London: Royal Microscopical Society,[1966]-c2004.).
35	Bolhuis, T., Pasop, J., Abelmann, L. & Lodder, J. in *AIP Conference Proceedings.* 271-278 (AIP).
36	Kundmann, M., Wilson, N., Torpy, A. & Zaluzec, N. MSA/MAS hyper-dimensional spectral file format-A straw-man proposal. *Microsc. microanal.* **17**, 860-861 (2011).
37	Maiden, A. M. & Rodenburg, J. M. An improved ptychographical phase retrieval algorithm for diffractive imaging. *Ultramicroscopy* **109**, 1256-1262 (2009).
38	Maiden, A., Humphry, M., Sarahan, M., Kraus, B. & Rodenburg, J. An annealing algorithm to correct positioning errors in ptychography. *Ultramicroscopy* **120**, 64-72 (2012).
39	Pennycook, T. J. *et al.* Efficient phase contrast imaging in STEM using a pixelated detector. Part 1: Experimental demonstration at atomic resolution. *Ultramicroscopy* **151**, 160-167 (2015).
40	Yang, H. *et al.* Electron ptychographic phase imaging of light elements in crystalline materials using Wigner distribution deconvolution. *Ultramicroscopy* **180**, 173-179 (2017).
41	de la Peña, F. *Advanced methods for Electron Energy Loss Spectroscopy core-loss analysis*, PhD thesis, Université de Paris-Sud, Paris, (2010).
42	Zaluzec, N. J. Evolution of Data Acquisition, Storage and Analysis in a Multi-user Facility. *Microsc. microanal.* **21**, 529-530 (2015).
43	Begum Gulsoy, Laura Bartola & attendees, a. o. in *2018 NIST/CHiMaD Microscopy Data Conference.*
44	Golub, G. H. & Reinsch, C. Singular value decomposition and least squares solutions. *Numerische mathematik* **14**, 403-420 (1970).

45  Hartigan, J. A. & Wong, M. A. Algorithm AS 136: A k-means clustering algorithm. *Journal of the Royal Statistical Society. Series C (Applied Statistics)* **28**, 100-108 (1979).
46  Stevens, A., Yang, H., Carin, L., Arslan, I. & Browning, N. D. The potential for Bayesian compressive sensing to significantly reduce electron dose in high-resolution STEM images. *Microscopy* **63**, 41-51 (2013).
47  Kovarik, L., Stevens, A., Liyu, A. & Browning, N. D. Implementing an accurate and rapid sparse sampling approach for low-dose atomic resolution STEM imaging. *Appl Phys Lett* **109**, 164102 (2016).
48  Könnecke, M. *et al.* The NeXus data format. *Journal of applied crystallography* **48**, 301-305 (2015).
49  Bicarregui, J., Matthews, B. & Schluenzen, F. J. S. R. N. PaNdata: Open Data Infrastructure for Photon and Neutron Sources. **28**, 30-35 (2015).
50  Kaufman, J. G. & Begley, E. F. MatML: A data interchange markup language. *Advanced Materials and Processes* **161** (2003).
51  De Carlo, F. *et al.* Scientific data exchange: a schema for HDF5-based storage of raw and analyzed data. **21**, 1224-1230 (2014).
52  Goldberg, I. G. *et al.* The Open Microscopy Environment (OME) Data Model and XML file: open tools for informatics and quantitative analysis in biological imaging. **6**, R47 (2005).
53  Ophus, P. *Electron Microscopy Datasets*, <https://emdatasets.com/format/> (2016).
54  Jackson, M., Simmons, J. & De Graef, M. MXA: a customizable HDF5-based data format for multi-dimensional data sets. *Modelling and Simulation in Materials Science and Engineering* **18**, 065008 (2010).
55  Groeber, M. A. & Jackson, M. A. DREAM. 3D: a digital representation environment for the analysis of microstructure in 3D. *Integrating Materials and Manufacturing Innovation* **3**, 5 (2014).
56  Somnath, S., Belianinov, A., Kalinin, S. V. & Jesse, S. Rapid mapping of polarization switching through complete information acquisition. *Nat Commun* **7**, 13290, doi:10.1038/ncomms13290 http://www.nature.com/articles/ncomms13290#supplementary-information (2016).
57  HDFView (The HDF Group, University of Illinois, Urbana-Champaign, Urbana, IL, USA, 2007).
58  Matlab, U. s. G. The mathworks. *Inc., Natick, MA* **1992** (1760).
59  Van Rossum, G. & Drake Jr, F. L. *Python reference manual*.  (Centrum voor Wiskunde en Informatica Amsterdam, 1995).
60  Kernighan, B. W. & Ritchie, D. M. *The C programming language*.  (2006).
61  Stroustrup, B. *The C++ programming language*.  (Pearson Education India, 2000).
62  Hejlsberg, A., Wiltamuth, S. & Golde, P. *C# language specification*.  (Addison-Wesley Longman Publishing Co., Inc., 2003).
63  Arnold, K., Gosling, J. & Holmes, D. *The Java programming language*.  (Addison Wesley Professional, 2005).
64  Wall, L., Christiansen, T. & Schwartz, R. L. Programming perl.  (1999).
65  Prabhat *et al.* ExaHDF5: An I/O Platform for Exascale Data Models, Analysis and Performance. *SciDAC 2011*.
66  Bray, T., Paoli, J., Sperberg-McQueen, C. M., Maler, E. & Yergeau, F. Extensible Markup Language (XML). *World Wide Web Journal* **2**, 27-66 (1997).
67  Kwon, T., Dhruv, N., Patwardhan, S. & Kwon, E. Common data format archiving of large-scale intelligent transportation systems data for efficient storage, retrieval, and portability. *Transportation Research Record: Journal of the Transportation Research Board*, 111-117 (2003).
68  Rew, R. & Davis, G. NetCDF: an interface for scientific data access. *IEEE computer graphics and applications* **10**, 76-82 (1990).
69  Folk, M., Heber, G., Koziol, Q., Pourmal, E. & Robinson, D. in *Proceedings of the EDBT/ICDT 2011 Workshop on Array Databases.*  36-47 (ACM).

70	Francesc Alted, M. D., Stephan Hoyer, John Kirkham, Alistair Miles, Mamy Ratsimbazafy, Matthew Rocklin, Vincent Schut, Anthony Scopatz, Prakhar Goel.    (GitHub, 2018).
71	Maia, F. R. J. N. m. The coherent X-ray imaging data bank.  **9**, 854 (2012).
72	Davison, A. P., Mattioni, M., Samarkanov, D. & Telenczuk, B. Sumatra: a toolkit for reproducible research. *Implementing reproducible research* **57** (2014).
73	Hinsen, K. ActivePapers: a platform for publishing and archiving computer-aided research. *F1000Research* **3** (2014).
74	Fomel, S., Sava, P., Vlad, I., Liu, Y. & Bashkardin, V. Madagascar: Open-source software project for multidimensional data analysis and reproducible computational experiments. *Journal of Open Research Software* **1** (2013).
75	Oxvig, C. S., Arildsen, T. & Larsen, T. in *Proceedings of the 15th Python in Science Conference.*  11-17.
76	Wilkinson, M. D. *et al.* The FAIR Guiding Principles for scientific data management and stewardship. *Scientific data* **3** (2016).
77	*Statement on Digital Data Management*, <https://science.energy.gov/funding-opportunities/digital-data-management/> (2016).
78	Nečas, D. & Klapetek, P. Gwyddion: an open-source software for SPM data analysis. *Open Physics* **10**, 181-188 (2012).
79	Cueva, P., Hovden, R. & Muller, D. Cornell Spectrum Imager: Open Source Spectrum Analysis with ImageJ. *Microsc. microanal.* **17**, 792-793 (2011).
80	VanderPlas, J. in *PyCon 2017.*
81	Barua, A., Thomas, S. W. & Hassan, A. E. What are developers talking about? an analysis of topics and trends in stack overflow. *Empirical Software Engineering* **19**, 619-654 (2014).
82	Walt, S. v. d., Colbert, S. C. & Varoquaux, G. The NumPy array: a structure for efficient numerical computation. *Computing in Science & Engineering* **13**, 22-30 (2011).
83	Hoyer, S. & Hamman, J. xarray: ND labeled Arrays and Datasets in Python. *Journal of Open Research Software* **5** (2017).
84	Behnel, S. *et al.* Cython: The best of both worlds. *Computing in Science & Engineering* **13**, 31-39 (2011).
85	Van der Walt, S. *et al.* scikit-image: image processing in Python. *PeerJ* **2**, e453 (2014).
86	Bradski, G. & Kaehler, A. OpenCV. *Dr. Dobb's journal of software tools* **3** (2000).
87	Jones, E., Oliphant, T. & Peterson, P. {SciPy}: open source scientific tools for {Python}.  (2014).
88	Pedregosa, F. *et al.* Scikit-learn: Machine learning in Python. *Journal of machine learning research* **12**, 2825-2830 (2011).
89	Chollet, F. Keras: The python deep learning library. *Astrophysics Source Code Library* (2018).
90	Abadi, M. *et al.* in *OSDI.*  265-283.
91	Paszke, A., Gross, S., Chintala, S. & Chanan, G.    (2017).
92	Collette, A. *Python and HDF5: Unlocking Scientific Data*.  (" O'Reilly Media, Inc.", 2013).
93	Summerfield, M. *Rapid GUI Programming with Python and Qt: The Definitive Guide to PyQt Programming (paperback)*.  (Pearson Education, 2007).
94	Klöckner, A. *et al.* PyCUDA and PyOpenCL: A scripting-based approach to GPU run-time code generation. *Parallel Computing* **38**, 157-174 (2012).
95	Sergeev, A. & Del Balso, M. Horovod: fast and easy distributed deep learning in TensorFlow. *arXiv preprint arXiv:1802.05799* (2018).
96	Rocklin, M. in *Proceedings of the 14th Python in Science Conference.*   (Citeseer).
97	Dalcin, L.    (2013).
98	Price-Whelan, A. M. & Foreman-Mackey, D. schwimmbad: A uniform interface to parallel processing pools in Python. *The Journal of Open Source Software* **2** (2017).
99	Hunter, J. D. Matplotlib: A 2D graphics environment. *Computing in science & engineering* **9**, 90-95 (2007).
100	VanderPlas, J. in *PyCon 2017.*

101    Stern, B. A. in *Space 2000*     1011-1015 (2000).
102    Oxvig, C., Pedersen, P., Arildsen, T., Østergaard, J. & Larsen, T. Magni: A python package for compressive sampling and reconstruction of atomic force microscopy images. *Journal of Open Research Software* **2** (2014).
103    Jesse, S. & Kalinin, S. V. Band excitation in scanning probe microscopy: sines of change. *J Phys D Appl Phys* **44**, doi:Artn 464006
Doi 10.1088/0022-3727/44/46/464006 (2011).
104    Jesse, S. *et al.* Band Excitation in Scanning Probe Microscopy: Recognition and Functional Imaging. *Annu Rev Phys Chem* **65**, 519-536, doi:DOI 10.1146/annurev-physchem-040513-103609 (2014).
105    Somnath, S., Kalinin, S. V. & Jesse, S. G-mode-Full Information Capture Applied to Scanning Probe Microscopy. *Microsc. microanal.* **23**, 184-185 (2017).
106    Jesse, S., Collins, L., Neumayer, S., Somnath, S. & Kalinin, S. V. in *Kelvin Probe Force Microscopy*     49-99 (Springer, 2018).
107    Belianinov, A., Kalinin, S. V. & Jesse, S. Complete information acquisition in dynamic force microscopy. *Nat Commun* **6**, doi:ARTN 6550
DOI 10.1038/ncomms7550 (2015).
108    Kluyver, T. *et al.* in *ELPUB.*  87-90.
109    ipywidgets: Interactive HTML widgets for Jupyter notebooks and the IPython kernel. v. 7.4.2 (Python Package Index (pypi), 2015).
110    Kurtzer, G. M., Sochat, V. & Bauer, M. W. Singularity: Scientific containers for mobility of compute. *PloS one* **12**, e0177459 (2017).
111    Boettiger, C. An introduction to Docker for reproducible research. *ACM SIGOPS Operating Systems Review* **49**, 71-79 (2015).
112    Jesse, S. *et al.* Big data analytics for scanning transmission electron microscopy ptychography. *Scientific reports* **6**, 26348 (2016).
113    Somnath, S. *et al.* Feature extraction via similarity search: application to atom finding and denoising in electron and scanning probe microscopy imaging. *Adv Struct Chem Imag* **4**, 3 (2018).
114    Vasudevan, R. K. *et al.* Big data in reciprocal space: Sliding fast Fourier transforms for determining periodicity. *Appl Phys Lett* **106**, 091601 (2015).
115    Sincraian, P. R. *Pycroscopy downloads statistics*, <http://pepy.tech/project/pycroscopy> (2018).
116    Sincraian, P. R. *pyUSID downloads statistics*, <http://pepy.tech/project/pyUSID> (2018).
117    Loeliger, J. & McCullough, M. *Version Control with Git: Powerful tools and techniques for collaborative software development*.  (" O'Reilly Media, Inc.", 2012).
118    Spinellis, D. Version control systems. *IEEE Software* **22**, 108-109 (2005).
119    Fowler, M. & Foemmel, M. Continuous integration. *Thought-Works)* http://www. thoughtworks. com/Continuous Integration. pdf* **122**, 14 (2006).
120    Travis CI - test and deploy your code with confidence (Travis CI, GmbH, Berlin, Germany, 2011).
121    CricleCI - Automate your development process quickly, safely, and at scale. (CricleCI, San Fransisco, CA, 2011).
122    Continuous Integration solution for Windows and Linux (AppVeyor, Vancouver, BC, Canada, 2011).
123    Vasudevan, R. K., Jesse, S., Smith, C. R. & Somnath, S. in *2018 Center for Nanophase Materials Science User Meeting.*
124    Vasudevan, R. K., Jesse, S. & Ziatdinov, M. in *2018 Center for Nanophase Materials Science User Meeting.*
125    Jesse, S. *et al.* in *Microscopy and Microanalysis 2018 Meeting.*
126    Jesse, S., Belianinov, A., Somnath, S. & Smith, C. R. in *Materials Research Society Fall 2017 Meeting.*

127     A. Gilad Kusne, Alexei Belianinov, Daniel Samarov & Takeuchi, I. in *Machine Learning for Materials Research Bootcamp 2018 & Workshop on Machine Learning Quantum Materials.*
128     Wilson, G. *et al.* Best practices for scientific computing. **12**, e1001745 (2014).
129     Allcock, W. *et al.* in *Proceedings of the 2005 ACM/IEEE conference on Supercomputing.*  54 (IEEE Computer Society).
130     Gray, J. *et al.* Scientific data management in the coming decade. *Acm Sigmod Record* **34**, 34-41 (2005).
131     Fernández, L. *et al.* Jupyterhub at the ESS. An Interactive Python Computing Environment for Scientists and Engineers.  (2016).
132     Blaiszik, B. *et al.* The Materials Data Facility: Data services to advance materials science research. *JOM* **68**, 2045-2052 (2016).
133     Caffaro, J. & Kaplun, S. Invenio: A modern digital library for grey literature. (2010).
134     Winn, J. Open data and the academy: an evaluation of CKAN for research data management. (2013).
135     Pandolfi, R. J. *et al.* Xi-cam: a versatile interface for data visualization and analysis. *Journal of synchrotron radiation* **25** (2018).
136     Lee, J.-G. & Kang, M. Geospatial big data: challenges and opportunities. *Big Data Research* **2**, 74-81 (2015).
137     Noy, N. *Making it easier to discover datasets*, <https://www.blog.google/products/search/making-it-easier-discover-datasets/> (2018).
138     de la Peña, F. *et al.* Electron microscopy (Big and Small) data analysis with the open source software package HyperSpy. *Microsc. microanal.* **23**, 214-215 (2017).
139     Clausen, A.   (ed Deiter Weber) (GitHub, 2018).
140     Lingerfelt, E. J. *et al.* BEAM: A Computational Workflow System for Managing and Modeling Material Characterization Data in HPC Environments. *Procedia Computer Science* **80**, 2276-2280, doi:http://dx.doi.org/10.1016/j.procs.2016.05.410 (2016).
141     Lam, S. K., Pitrou, A. & Seibert, S. in *Proceedings of the Second Workshop on the LLVM Compiler Infrastructure in HPC.*  7 (ACM).